\documentclass[prl,preprint,endfloats]{revtex4}

\usepackage[driver=dvipdfm,letterpaper]{geometry}

\usepackage{graphicx}
\usepackage{bm}
\usepackage{pifont}
\usepackage{amsmath}

\begin{document}

\title{Nonreciprocal spin wave propagation in chiral-lattice ferromagnets}

\author{S. Seki$^{1,2}$}
\author{Y. Okamura$^{3}$} 
\author{K. Kondou$^{1}$}
\author{K. Shibata$^{3}$}
\author{M. Kubota$^{1,4}$}
\altaffiliation{Present Address: Technology and Business Development Unit, Murata Manufacturing Co., Ltd., Nagaokakyo, Kyoto 617-8555, Japan}
\author{R. Takagi$^{1}$} 
\author{F. Kagawa$^{1}$}
\author{M. Kawasaki$^{1,3}$}
\author{G. Tatara$^{1}$}
\author{Y. Otani$^{1,5}$}
\author{Y. Tokura$^{1,3}$} 
\affiliation{$^1$ RIKEN Center for Emergent Matter Science (CEMS), Wako 351-0198, Japan}
\affiliation{$^2$ PRESTO, Japan Science and Technology Agency (JST), Tokyo 113-8656, Japan}
\affiliation{$^3$ Department of Applied Physics and Quantum Phase Electronics Center (QPEC), University of Tokyo, Tokyo 113-8656, Japan}
\affiliation{$^4$ Research and Development Headquarters, ROHM Co., Ltd., Kyoto 615-8585, Japan}
\affiliation{$^5$ Institute for Solid State Physics, University of Tokyo, Kashiwa 277-8581, Japan}

\date{}

\begin{abstract}
Spin current, i.e. the flow of spin angular momentum or magnetic moment, has recently attracted much attention as the promising alternative for charge current with better energy efficiency. Genuine spin current is generally carried by the spin wave (propagating spin precession) in insulating ferromagnets, and should hold the chiral symmetry when it propagates along the spin direction. Here, we experimentally demonstrate that such a spin wave spin current (SWSC) shows nonreciprocal propagation characters in a chiral-lattice ferromagnet. This phenomenon originates from the interference of chirality between the SWSC and crystal-lattice, which is mediated by the relativistic spin-orbit interaction. The present finding enables the design of perfect spin current diode, and highlights the importance of the chiral aspect in SWSC.

\end{abstract}

\maketitle

\begin{figure}[htbp]
\begin{center}
\includegraphics*[width=14cm]{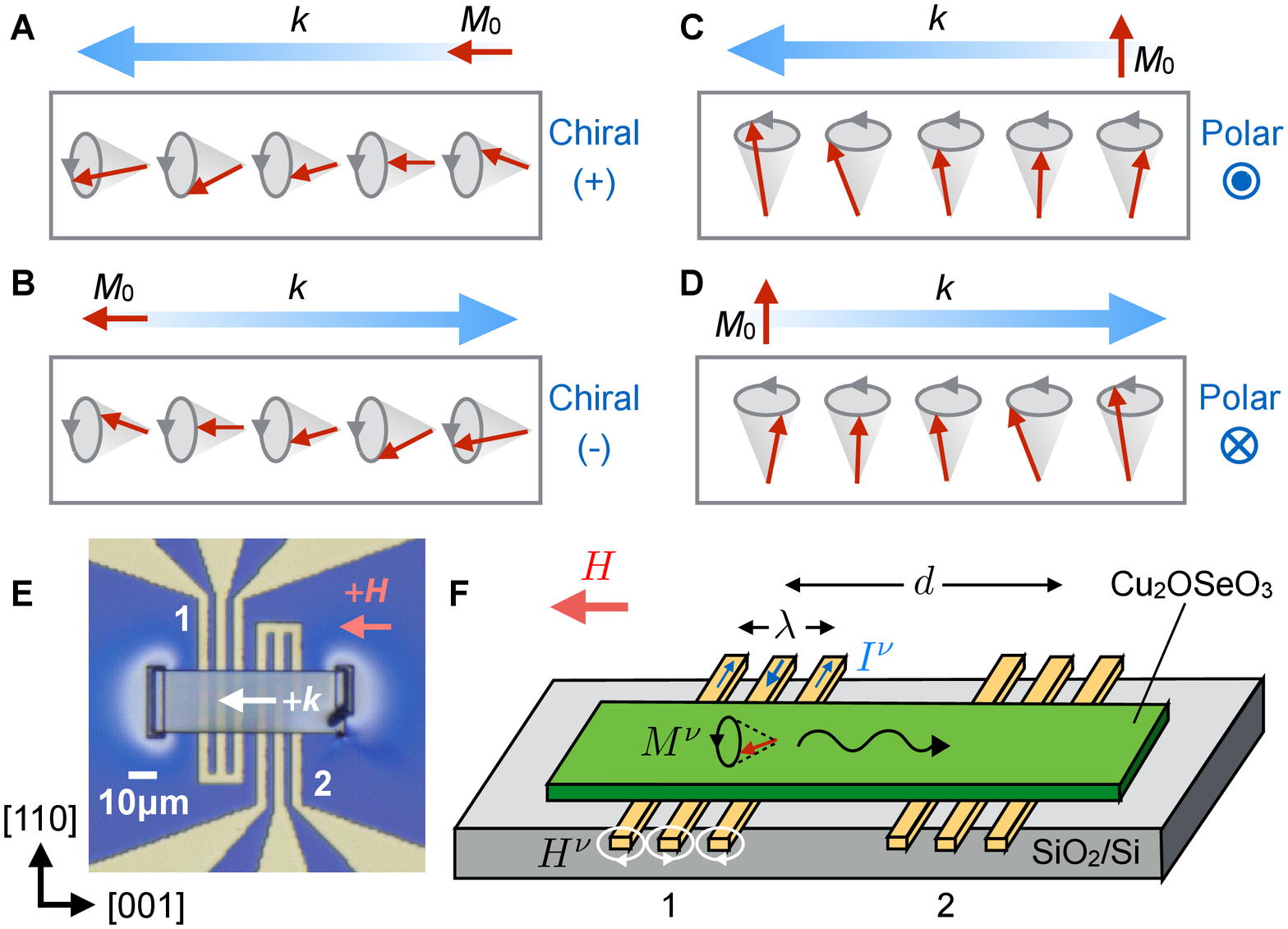}
\caption{(A)-(D) Spin waves for the uniform collinear ferromagnetic state, characterized by various combinations of the wave vector $\vec{k}$ and the uniform magnetization component $\vec{M}_0$, which can be considered as the flow of magnetic moment (i.e. spin current). Such a spin wave spin current belongs to chiral (polar) symmetry for the $\vec{k} \parallel \vec{M}_0$ ($\vec{k} \perp \vec{M}_0$) configuration, and the reversal of $\vec{k}$ gives opposite sign of chirality (polarity). The red and gray arrows represent the directions of local magnetization and its precession, respectively. (E) The optical microscope image and (F) schematic illustration of the device structure used for spin wave spectroscopy. In (E), the directions for positive sign of $\vec{H}$ and $\vec{k}$ are also indicated.}
\label{FigScheme}
\end{center}
\end{figure}

Electron is a particle characterized by the charge and spin degree of freedom. In contrast to the charge current accompanied by Joule heat loss, spin current can be dissipationless and potentially minimize the energy consumption associated with the information processing\cite{SC_Review, SpinCurrentMurakami, SpinHallExp, SpinHallTheory}. Spin current is generally carried by the spin-polarized conduction electrons in metallic system, as well as the spin wave in insulating system\cite{YIG_SP, YIG_SSE}. In particular, the latter spin wave spin current (SWSC) possesses many advantages over the former one, since it can avoid the simultaneous flow of charge current and has much longer propagation length. From the viewpoint of the symmetry, there are two types of SWSC for ferromagnets, depending on the directional relationship between the carried magnetic moment $\vec{M}_0$ (parallel to the external magnetic field $\vec{H}$) and the wave vector $\vec{k}$. In case of $\vec{M}_0 \parallel \vec{k}$ (Fig. \ref{FigScheme}A), the SWSC doesn't have any mirror plane or space-inversion center, and belongs to the chiral symmetry\cite{SW_Symmetry, comment}. In contrast, the SWSC with $\vec{M}_0 \perp \vec{k}$ configuration (Fig. \ref{FigScheme}C) has the polar symmetry with the polar axis normal to both $\vec{M}_0$ and $\vec{k}$. For each case, the reversal of $\vec{k}$ gives the SWSC with opposite chirality or polarity (Fig. \ref{FigScheme}, B and D).

The above analysis predicts that the SWSCs propagating along the positive and negative direction can show different propagation characters, when placed in the chiral or polar environment depending on the symmetry of SWSC. For example, the surface (or interface) is always characterized by the structural polarization normal to the surface. When both $\vec{M}_0 (\parallel \vec{H})$ and $\vec{k}$ are confined within the surface plane keeping the $\vec{M}_0 \perp \vec{k}$ relationship, the polar axis of SWSC (Fig. \ref{FigScheme}, C and D) becomes parallel or antiparallel to the polarization of the surface and thus the asymmetric spin wave propagation between $\pm k$ can be expected. Such a surface-induced spin wave nonreciprocity has first been predicted by Damon and Eshbach more than 50 years ago\cite{DEFirst}, and then verified by various experimental techniques such as spin wave spectroscopy\cite{Surface_YIG, MSW_Text}, spin-polarized electron energy loss spectroscopy\cite{SW_DM_SPEEL}, thermography\cite{UniHeat}, and Brillouin light scattering\cite{SW_DM_BLS}. In particular, the $k \rightarrow 0$ mode mediated by the magnetic dipole-dipole interaction in this configuration is called magnetostatic surface wave (representing that it can propagate only at the surface of the sample), and is known for its unidirectional propagation character\cite{DEFirst, MSW_Text, MSW_Text2}. Nevertheless, such a surface-driven nonreciprocity  mostly cancels out as the entire sample, since the sign of polarity is generally opposite between the top and bottom surfaces.

In contrast, the interplay between the SWSC and the chiral medium has hardly been investigated. One reason is that the coexistence of crystallographic chirality and ferromagnetism is very rare in the real materials. Recently, however, some metallic ferromagnets with the chiral B20 crystallographic lattice (such as MnSi, Fe$_{1-x}$Co$_x$Si, and FeGe) have been found to host magnetic skyrmions, i.e. nanometer-sized vortex-like swirling spin texture with particle nature\cite{SkXTheory, NeutronMnSi, TEMFeCoSi, SkXReviewTokura}. Skyrmions are now attracting much attention as the potential information carrier for the high-density magnetic storage device\cite{SkXReviewTokura, SkXReviewFert}, and the above finding has promoted further search of similar chiral-lattice ferromagnetic materials. Soon after their discovery in the B20 materials, it has been reported that a chiral-lattice ferromagnetic insulator Cu$_2$OSeO$_3$ (with cubic space group $P2_13$) can also host magnetic skyrmions, for the narrow temperature region just below $T_c \sim 59$ K\cite{Cu2OSeO3_Seki, Cu2OSeO3_SANS_Seki, Cu2OSeO3_SANS_Pfleiderer, Onose}. Notably, this material offers an ideal opportunity to investigate the property of SWSC under the chiral environment. In this work, we have examined the nonreciprocal nature of spin wave propagation for this chiral-lattice magnet in terms of spin wave spectroscopy.

The basic concept of the spin wave spectroscopy\cite{AESWS_doppler_Science, AESWS_doppler_B} as well as the employed device structure is summarized in Fig. \ref{FigScheme}, E and F. A pair of Au coplanar waveguides (ports 1 and 2) were fabricated on the oxidized silicon substrate, and the plate-shaped single crystal of Cu$_2$OSeO$_3$ was placed across them. Here, the chirality of each Cu$_2$OSeO$_3$ crystal (D or L) is checked by measuring the natural optical activity at light wavelength 1310 nm in advance. When the oscillating electric current $I^\nu$ of gigahertz frequency $\nu$ is injected into one of the waveguides, $I^\nu$ generates oscillating magnetic field $H^\nu$ and excites spin wave (i.e. coherent magnetization oscillation $M^\nu$) in the Cu$_2$OSeO$_3$ sample. The propagating spin wave causes an additional magnetic flux on the waveguides, and induces the oscillating electric voltage $V^\nu$ following the Faraday's law. By measuring the spectrum of complex inductance $L_{nm} (\nu)$ as defined by $V^\nu_n = \sum_m L_{nm}(\nu) \frac{dI^\nu_m}{dt}$ (with $m$ and $n$ representing the port numbers used for the excitation and detection, respectively) with the vector network analyzer (VNA), we can evaluate both magnitude and phase of propagating spin wave. The spin wave contribution to the inductance spectrum $\Delta L_{nm} (\nu) = L_{nm} (\nu) - L^\textrm{ref}_{nm}(\nu)$ is derived by the subtraction of the common background $L^\textrm{ref}_{nm}(\nu)$ from the raw data $L_{nm} (\nu)$. Here, $L_{nm} (\nu)$ taken at $H=2650$ Oe is adopted as $L^\textrm{ref}_{nm}(\nu)$, where the magnetic resonance is absent within our target frequency range from 2GHz to 7GHz. The wave number $k$ of excited spin wave is determined by the spatial periodicity $\lambda (= 12 \mu$m) of the waveguide pattern and the associated current density $I^\nu (x)$\cite{AESWS_doppler_Science, AESWS_doppler_B}. Its Fourier transform $|\tilde{I}^\nu (k)|^2$ has the main peak at $k_p = 0.50 \mu$m$^{-1}$ with the full width at half maximum (FWHM) of $\delta k = 0.37 \mu$m$^{-1}$ as plotted in Fig. \ref{FigDisp}C, satisfying the relationship $k_p \sim 2\pi/\lambda$. To investigate the property of SWSC of the chiral symmetry (Fig. \ref{FigScheme}, A and B), the $H \parallel k \parallel [001]$ configuration is always adopted here. In this setup, the $k \rightarrow 0$ mode is called magnetostatic backward volume wave, which propagates through the entire volume of the sample with the negative group velocity\cite{DEFirst, MSW_Text, MSW_Text2}. In the centrosymmetric materials, this mode should not show any nonreciprocal propagation nature.

\begin{figure}
\begin{center}
\includegraphics*[width=15cm]{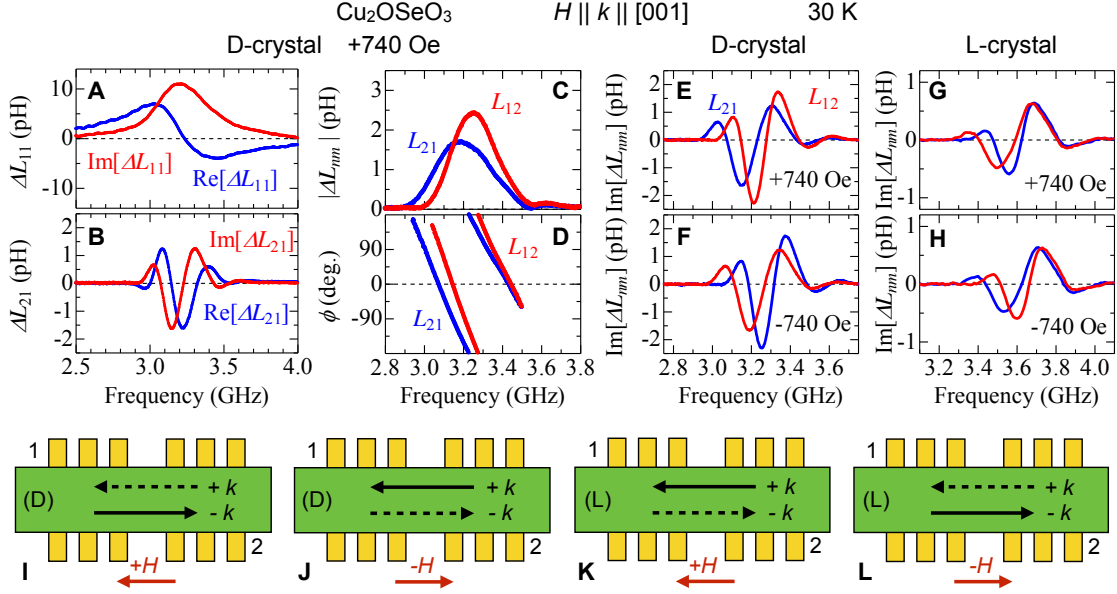}
\caption{(A)-(H) Spin wave contribution to inductance spectrum $\Delta L_{nm} = $Re$[\Delta L_{nm}] + i$ Im$[\Delta L_{nm}]$ = $|\Delta L_{nm}| \exp[i\phi]$, measured for the D- or L-chirality of Cu$_2$OSeO$_3$ single crystal with the $H \parallel k \parallel [001]$ configuration at 30 K. All the data are taken at the uniform collinear ferromagnetic state. (A) and (B) Real and imaginary part of self inductance $\Delta L_{11}$ and mutual inductance $\Delta L_{21}$ measured for the D-crystal at $H = +740$ Oe. For the same configuration, (C) magnitude $|\Delta L_{nm}|$ and (D) phase $\phi$ of $\Delta L_{21}$ and $\Delta L_{12}$ are also plotted. (E)-(H) Imaginary part of $\Delta L_{21}$ and $\Delta L_{12}$, measured with various combinations of magnetic field direction ($H=\pm 740$ Oe) and crystallographic chirality (D or L). Note that the deviation of the overall signal magnitude and resonance frequency between D- and L-crystal is due to the slight difference in their sample size and associated demagnetizing field. The corresponding experimental configurations as well as the expected sign of nonreciprocity are summarized in (I)-(L). Here, the spin wave characterized by the wave vector $+ k$ ($-k$) contributes to $\Delta L_{21}$ ($\Delta L_{12}$), and the solid and dashed arrows represent the different propagation characters.}
\label{FigSpectra}
\end{center}
\end{figure}

First, we have investigated the nature of SWSC in the uniform collinear ferromagnetic state with saturated magnetization. Figure \ref{FigSpectra}, A and B indicate the real and imaginary part of $\Delta L_{11}$ and $\Delta L_{21}$ spectra measured at +740 Oe, i.e. in the collinear ferromagnetic state, for the D-chirality of the Cu$_2$OSeO$_3$ crystal. The self-inductance $\Delta L_{11}$ represents the efficiency of the local spin wave excitation, and the ferromagnetic resonance characterized by Lorentzian shape of spectrum can be identified at 3.2 GHz. In contrast, the mutual-inductance $\Delta L_{21}$ reflects the propagation character of spin wave between the two waveguides, and the finite oscillating signal can be detected around the same resonance frequency. Hereafter, we focus on the comparison between $\Delta L_{21}$ and $\Delta L_{12}$, each of which stands for the propagating spin wave characterized by the wave vector $+k$ and $-k$, respectively. To interpret the data more intuitively, the spectra of $|\Delta L_{nm}|$ and $\phi$ as defined with $\Delta L_{nm} =$ Re$[\Delta L_{nm}] + i$ Im$[\Delta L_{nm}] = |\Delta L_{nm}| \exp[i\phi]$ are plotted in Fig. \ref{FigSpectra}, C and D. $|\Delta L_{21}|$ and $|\Delta L_{12}|$ express the magnitude of spin wave after the propagation along the positive and negative directions, and both spectra show a peak structure. Notably, the peak frequency $\nu_p$ as well as the peak intensity $|\Delta L^p_{nm}|$ are clearly different between $\pm k$. On the other hand, $\phi$ represents the phase delay of spin wave for a transmission between the two waveguides separated by the distance $d (= 20 \mu$m), and will satisfy the relationship $\phi = kd$ when a single spin wave mode is assumed\cite{AESWS_doppler_Science, AESWS_doppler_B}. This means that the $\phi$ spectrum directly reflects the spin wave dispersion relationship, and its slope gives the group velocity $v_g = \frac{\partial \omega}{\partial k} = 2\pi d (\frac{\partial \phi}{\partial \nu})^{-1}$. The clear difference in the $\phi$ slope between the ones derived from $\Delta L_{21}$ and $\Delta L_{12}$ indicates that $v_g$ of spin wave is not equal between $\pm k$. The above results establish that the propagation character of spin wave in this configuration is nonreciprocal, from both aspects of magnitude and group velocity.

To reveal the origin of nonreciprocity, the measurements of Im[$\Delta L_{21}$] and Im[$\Delta L_{12}$] were performed with various combinations of the magnetic field direction ($H = \pm 740$ Oe) and crystallographic chirality (D and L) for Cu$_2$OSeO$_3$ (Fig. \ref{FigSpectra}, E to H). Figure \ref{FigSpectra}E shows the data for the D-crystal at $+740$ Oe. The spectra of Im[$\Delta L_{21}$] and Im[$\Delta L_{12}$] are characterized by the signal oscillating with different period and magnitude, in agreement with the feature observed for $|\Delta L_{nm}|$ and $\phi$. For the opposite sign of applied $H$ (Fig. \ref{FigSpectra}F), the spectral shapes for Im[$\Delta L_{21}$] and Im[$\Delta L_{12}$] (i.e. the sign of nonreciprocity) are reversed. Likewise, the employment of opposite chirality of crystal (i.e. L-crystal) also reverses the sign of nonreciprocity (Fig. \ref{FigSpectra}, G and H). In Fig. \ref{FigSpectra}, I to L, the symmetrically expected sign of nonreciprocity for each experimental configuration is summarized. In general, the SWSC is expressed as the product of magnetic moment $\vec{M}_0$ ($\parallel \vec{H}$) and wave vector $\vec{k}$ (Fig. \ref{Hamiltonian}, A and B). This means that $H$-reversal has the same effect as $k$-reversal, and thus results in the opposite sign of nonreciprocity. On the other hand, the application of space-inversion operation to the system reverses both the crystallographic chirality and the $\vec{k}$-direction keeping $\vec{H}$ and $\vec{M}_0$ unchanged, indicating that the sign of nonreciprocity should be opposite between L-crystal and D-crystal. The above symmetry-based analysis is in agreement with the experimental results, which proves that the observed spin wave nonreciprocity originates from the chiral nature of crystal lattice.

\begin{figure*}
\begin{center}
\includegraphics*[width=7cm]{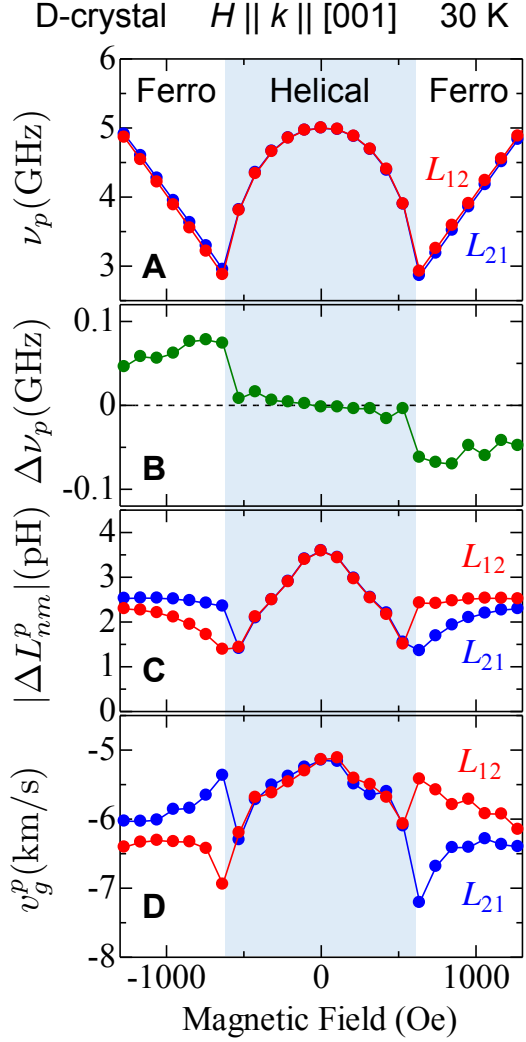}
\caption{Magnetic field dependence of spin wave nonreciprocity between $\Delta L_{21}$ and $\Delta L_{12}$ (i.e. $+k$ and $-k$), measured for the D-crystal of Cu$_2$OSeO$_3$ with the $H \parallel k \parallel [001]$ configuration at 30 K. (A) and (B) indicate the magnetic resonance frequencies $\nu_p$ giving the peak value of $|\Delta L_{nm}|$, and their difference between $\pm k$ (i.e. $\Delta \nu_p = \nu_p(+k)-\nu_p(-k)$), respectively. In (C) and (D), the corresponding peak value $|\Delta L^p_{nm}|$ and the group velocity $v^p_g$ at the frequency $\nu_p$ are also plotted.}
\label{FigHdep}
\end{center}
\end{figure*}

\begin{figure}
\begin{center}
\includegraphics*[width=12cm]{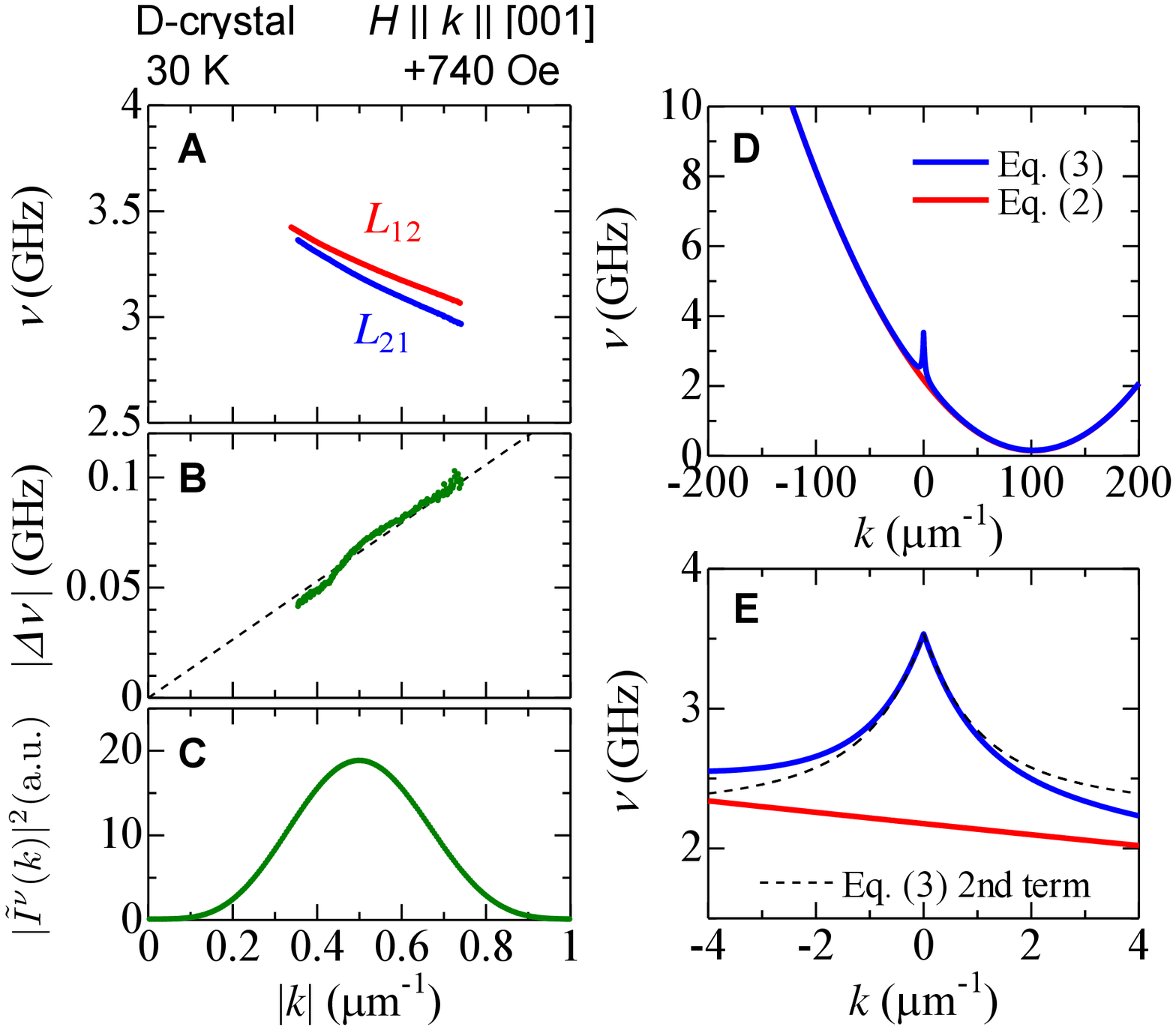}
\caption{(A) Spin wave dispersion for the D-crystal of Cu$_2$OSeO$_3$ at $H = + 740$ Oe (i.e. collinear ferromagnetic state), experimentally deduced by analyzing the $\phi$ spectrum in Fig. \ref{FigSpectra}D. The one obtained from $\Delta L_{21}$ ($\Delta L_{12}$) corresponds to positive (negative) $k$-value, and the frequency difference between $\pm k$ (i.e. $\Delta \nu (|k|) = \nu(+|k|)-\nu(-|k|)$) is also plotted in (B). (C) Wave number distribution of excitation current $\tilde{I}^\nu (k)$, obtained by the Fourier transform of the wave guide pattern. (D) and (E) Spin wave dispersions for the collinear ferromagnetic state calculated based on Eq. \ref{DispWithDipole} or Eq. \ref{DispNoDipole}, the latter of which ignores the effect of magnetic dipole-dipole interaction. The black dashed line represents the contribution of the second term in Eq. \ref{DispWithDipole}. Here, the assumed material parameters are $l=2\mu$m, $D = 3.4 \times 10^{-4}$ J/m$^2$, $J = 3.4 \times 10^{-12}$ J/m, $K = 6.9 \times 10^{3}$ J/m$^3$, $\gamma/2\pi = 29$ GHz T$^{-1}$, $\mu_0 M_s = 0.12$ T, $\mu_0 H$ = 0.074 T, $V_0 = 89$ \AA$^3$, and $S=0.44$.} 
\label{FigDisp}
\end{center}
\end{figure}

Next, we investigated the magnetic field dependence of nonreciprocity. In Fig. \ref{FigHdep}, A and B, the peak frequency $\nu_p$ for $|\Delta L_{21}|$ and $|\Delta L_{12}|$ (defined as $\nu_p (+k)$ and $\nu_p (-k)$), as well as the difference between them $(\Delta \nu_p = \nu_p (+k) - \nu_p (-k))$, are plotted as a function of $H$. Cu$_2$OSeO$_3$ is known to host the helical spin order for $H=0$\cite{Cu2OSeO3_Seki, Cu2OSeO3_SANS_Seki, Cu2OSeO3_SANS_Pfleiderer}, while it is replaced with the uniform collinear ferromagnetic order for $H>600$ Oe. The magnetic resonance frequency is gradually suppressed by $H$ in the helical spin state, and then shows $H$-linear increase in the collinear ferromagnetic state. These behaviors are consistent with the previous reports\cite{Onose, B20_FMR, Okamura_First}. Notably, the magnitude of nonreciprocity is essentially dependent on the underlying magnetic structure. While the relatively large shift of resonance frequency $\Delta \nu_p \sim 0.05$ GHz between $\pm k$ is always observed for the collinear ferromagnetic state, such a nonreciprocity suddenly vanishes upon the transition into the helical spin state. Similar behavior is also observed for the peak intensity $|\Delta L^p_{nm}|$ (Fig. \ref{FigHdep}C) and group velocity $v^p_g$ (Fig. \ref{FigHdep}D) deduced at $\nu_p$. For all these properties, clear nonreciprocity is observed only in the collinear ferromagnetic state. Note that $\nu_p$, $|\Delta L^p_{nm}|$, and $v_g^p$ basically reflect the frequency, magnitude, and group velocity of spin wave for $|k|=k_p$. At maximum, the frequency shift $\Delta \nu_p = 0.07$ GHz and the change in $v_g^p$ ($|\Delta L^p_{nm}|$) up to 25 $\%$ (40$\%$) can be obtained between $\pm k$. Their sign of nonreciprocity is confirmed to reverse for $H$-reversal.

To clarify the microscopic origin of observed nonreciprocity, we attempt to estimate the spin wave dispersion for this material. According to Ref. \cite{Kataoka, MnSi_Disp_New}, the magnetic Hamiltonian for the ferromagnets with chiral cubic lattice symmetry under the continuum approximation can be written as
\begin{equation}
\mathcal{H} =  \int \left [ \frac{J}{2}  (\nabla \vec{S})^2 - D \vec{S} \cdot [\nabla \times \vec{S}]-\frac{K}{2} \sum_{i} S_i ^4 - \frac{\gamma \hbar}{V_0} \mu_0 \vec{H} \cdot \vec{S} \right ] d\vec{r} ,
\label{Hamiltonian}
\end{equation}
where $J$, $D$, and $K$ describe the magnitude of ferromagnetic exchange, Dzyaloshinskii-Moriya (DM), and cubic anisotropy term, respectively. $\vec{S}$ is dimensionless parameter representing the vector spin density. $\gamma$, $\mu_0$, $h=2\pi \hbar$, and $V_0$ are gyromagnetic ratio, vacuum magnetic permeability, Planck constant, and the volume of formula unit cell of Cu$_2$OSeO$_3$, respectively. For the $H \parallel k \parallel [001]$ configuration, the spin wave dispersion $\nu(k)$ for the uniform collinear ferromagnetic state is described as\cite{Kataoka}
\begin{equation}
\nu =  \frac{V_0}{h} [ 2DSk + JSk^2 + 2KS^3] + \frac{\gamma}{2\pi} \mu_0 H.
\label{DispNoDipole}
\end{equation}
In the real sample, this dispersion is further modified by the additional contribution of the magnetic dipole-dipole interaction, especially for the $k \rightarrow 0$ region. When the infinitely wide plate-shaped sample with the thickness $l$ is assumed and $H \parallel k \parallel [001]$ lies along the in-plane direction, Eq. \ref{DispNoDipole} can be rewritten as\cite{MSW_Text, MSW_Text2, MnSi_Disp_New}
\begin{equation}
\nu = \frac{2DSV_0 k}{h} + \frac{1}{h} \sqrt{(JSV_0 k^2 + 2KV_0 S^3 + \gamma \hbar \mu_0 H)(JSV_0 k^2 + 2KV_0 S^3 + \gamma \hbar \mu_0 H + \gamma \hbar \mu_0 M_s \frac{(1-e^{-|k|l})}{|k|l})},
\label{DispWithDipole}
\end{equation}
with $M_s$ being the saturation magnetization. In Fig. \ref{FigDisp}, D and E, the spin wave dispersion calculated based on Eq. \ref{DispWithDipole} with the material parameters estimated for Cu$_2$OSeO$_3$ is plotted. Equation \ref{DispWithDipole} can be approximated by Eq. \ref{DispNoDipole} except for the $k \rightarrow 0$ region, and gives parabolic dispersion with its minimum at $k=-D/J$. As $k$ approaches zero, however, the contribution of magnetic dipole-dipole interaction gradually increases the spin wave frequency. It causes the negative group velocity for the $k \rightarrow 0$ region, and this mode can be considered as a kind of magnetostatic backward volume wave\cite{MSW_Text, MSW_Text2}. Note that the first and second term in Eq. \ref{DispWithDipole} are odd and even functions of $k$, respectively, and thus only the former one proportional to $Dk$ can contribute to the spin wave nonreciprocity. This suggests that the observed nonreciprocity directly comes from the DM interaction, whose sign and magnitude reflect the chirality of the underlying crystallographic lattice through the relativistic spin-orbit interaction. 
  
Experimentally, the above spin wave dispersion relationship can be partly reproduced by analyzing the $\phi$ spectrum (Fig. \ref{FigSpectra}D). Given that the frequency $\nu = \nu_p$ corresponds to the wave number $k=k_p$ and the relationship $\phi = kd$ holds\cite{AESWS_doppler_Science, AESWS_doppler_B}, the dispersion relationship can be determined by $k=[\phi(\nu)-\phi(\nu_p)]/d+k_p$. In Fig. \ref{FigDisp}A, the spin wave dispersions $\nu(k)$ for the collinear ferromagnetic state deduced from the $\Delta L_{21}$ and $\Delta L_{12}$ spectra in Fig. \ref{FigSpectra}, C and D are plotted, each of which corresponds to the one for positive and negative $k$, respectively. For both cases, the concave-up dispersions with negative slopes are obtained, consistent with the prediction of Eq. \ref{DispWithDipole} for the $k \rightarrow 0$ region (Fig. \ref{FigDisp}E)\cite{MSW_Text, MSW_Text2, MnSi_Disp_New}. The dispersion curves for positive and negative $k$ show considerable deviation from each other, and their frequency shift $\Delta \nu (|k|) = \nu(+|k|) - \nu(-|k|)$ is plotted in Fig. \ref{FigDisp}B. $\Delta \nu$ is found to be almost proportional to $|k|$, which is consistent with the relationship $\Delta \nu = 4DSV_0 |k|/h$ expected from Eq. \ref{DispWithDipole}. The observed slope of $\Delta \nu$ gives $D \sim 5.5 \times 10^{-4}$ J/m$^2$, which roughly agrees with $D \sim 3.4 \times 10^{-4}$ J/m$^2$ estimated from $H$-dependence of magnetic resonance frequency\cite{B20_FMR}. These results firmly confirm that the observed spin wave nonreciprocity stems from the DM interaction associated with the chiral nature of crystallographic lattice. Note that for the helical spin state under the magnetic Hamiltonian given by Eq. \ref{Hamiltonian}, it has been proposed that the Brillouin zone is folded back with the helical spin modulation period due to the expansion of magnetic unit cell\cite{Kataoka, Helimagnon_Exp}. Such a folding back of magnon branch should extinguish the asymmetry between $\pm k$, which explains the observed disappearance of spin wave nonreciprocity in the helical spin state.

The above relationship $\Delta \nu \propto D|k|$ obtained for the ferromagnetic state suggests that $\Delta \nu$ linearly increases for larger $|k|$. Since the linewidth of the resonance peak in $\Delta L_{nm}$ spectrum is mainly determined by the $k$-distribution of waveguide for the present situation (see Supporting Online Material), the employment of the waveguide pattern with shorter wavelength and/or repeated meander shape\cite{AESWS_doppler_Science, AESWS_doppler_B} (i.e. enhancement of $k_p$ and suppression of $\delta k$) will completely resolve the frequency overlap of spin wave signal between $\pm k$. This means that for the given resonance frequency the SWSC can propagate along only one direction, and not along the opposite direction at all.
 
Recently, the relevance of DM interaction has also been discussed for the case of surface/interface-driven spin wave nonreciprocity\cite{SW_DM_Theory, SW_DM_Theory2, SW_DM_Theory3}, through the spin-polarized electron energy loss\cite{SW_DM_SPEEL} and Brillouin light scattering\cite{SW_DM_BLS} experiments. Their reported magnitudes of frequency shift between $\pm k$ are $\Delta \nu/|k| = 50$ MHz/$\mu$m$^{-1}$ for Pt/Co/Ni film\cite{SW_DM_BLS} and 240  MHz/$\mu$m$^{-1}$ for Fe double layer on W(110)\cite{SW_DM_SPEEL}, while the presently observed 130 MHz/$\mu$m$^{-1}$ for bulk Cu$_2$OSeO$_3$ is comparable with these systems. Note that the nonreciprocity in Cu$_2$OSeO$_3$ is for the volume spin wave and originates from the chiral lattice symmetry of the bulk crystal itself, unlike the conventional case of the surface/interface-driven nonreciprocity for the surface spin wave. Since the nonreciprocal volume spin wave can avoid the cancelation of nonreciprocity between the top and bottom surface inherent to the latter situation, our present finding offers a simple and promising route for the realization of perfect spin current diode with the 100$\%$ efficiency of rectification. It may also find unique spin caloritronic applications\cite{YIG_SSE, SpinCalo}, such as magnetically-tunable unidirectional heat conveyer\cite{UniHeat}.

From a broader perspective, any (quasi-)particle flow along the magnetic field direction should have the chiral symmetry\cite{SW_Symmetry}, and will show the similar nonreciprocal propagation character in the chiral-lattice compound. This has been experimentally confirmed for the light\cite{MChD_First, MChD_Helical} and conduction electron\cite{EMChA_First, EMChA_Ferro}, while their reported magnitude of nonreciprocity is generally very small. Our present observation of clear spin wave (or magnon) nonreciprocity in the chiral-lattice compound highlights the importance of the chiral aspect in SWSC.

\newpage

\section{Supporting Online Material}

\subsection{Experimental Details}

Single crystals of Cu$_2$OSeO$_3$ are grown by chemical vapor transport method\cite{COSO_First, COSO_Growth, Cu2OSeO3_Seki}. They are cut into the plate-like shape with widest faces parallel to the (110) plane, and polished with diamond slurry and colloidal silica. D- and L-chirality of crystals are distinguished by measuring the sign of natural optical activity at light wavelength $1310$ nm. Every piece of crystal shows optical rotation angle $\pm 16 ^\circ$/mm, and its single-domain nature is confirmed by the observation under polarized-light microscope. 

Our procedure of the measurement and data analysis associated with the spin wave spectroscopy basically follows the method proposed in Ref. \cite{AESWS_doppler_Science, AESWS_doppler_B}. A pair of coplanar waveguide patterns (ports 1 and 2) consisting of Au 195 nm / Ti 5 nm are deposited on a thermally oxidized silicon substrate using photo lithography and electron beam evaporation technique. The rectangular shape of Cu$_2$OSeO$_3$ crystals with typical size from 60 $\mu$m $\times$ 20 $\mu$m $\times$ 2 $\mu$m to 60 $\mu$m $\times$ 10 $\mu$m $\times$ 1 $\mu$m are extracted from the original bulk crystal pieces by focused ion beam micro-sampling technique, and placed across the waveguides with W deposition at both edges of the crystal. This device is put into the probe station equipped with GM refrigerator and horizontal electromagnet, and connected with a vector network analyzer (VNA) through the coaxial cable and GSG (ground-signal-ground) microprobe. The calibration is performed using Short-Open-Load-Throurgh coplanar standards. The spectrum of $S$-parameter ($S_{nm} (\nu, H)$ with $m$ and $n$ representing the port numbers used for the excitation and detection, respectively) is measured by VNA at various magnitudes of external magnetic field, and converted into the  impedance spectrum $Z_{nm} (\nu, H)$ assuming the characteristic impedance $Z_0 = 50 \Omega$\cite{StoZ}. The one at $H_\textrm{ref}=2650$ Oe is considered as the background, where no magnetic resonance appears within the target frequency range (2 GHz $\leq \nu \leq$ 7 GHz). From the subtraction of two impedance spectra, the spin wave contribution to the inductance spectrum $\Delta L_{nm} (\nu, H)$ is derived as $\Delta L_{nm} (\nu, H) = [Z_{nm} (\nu, H) - Z_{nm}(\nu, H_\textrm{ref})]/(i 2 \pi \nu)$. The input power into the waveguide is $-15$ dBm within the linear response regime, which is confirmed by measuring the power dependence of spectrum around the magnetic resonance frequency.  In this work, the $H \parallel k \parallel [001]$ configuration is always adopted.

\section{Design of coplaner waveguide and linewidth of magnetic resonance}

\begin{figure}
\begin{center}
\includegraphics*[width=14cm]{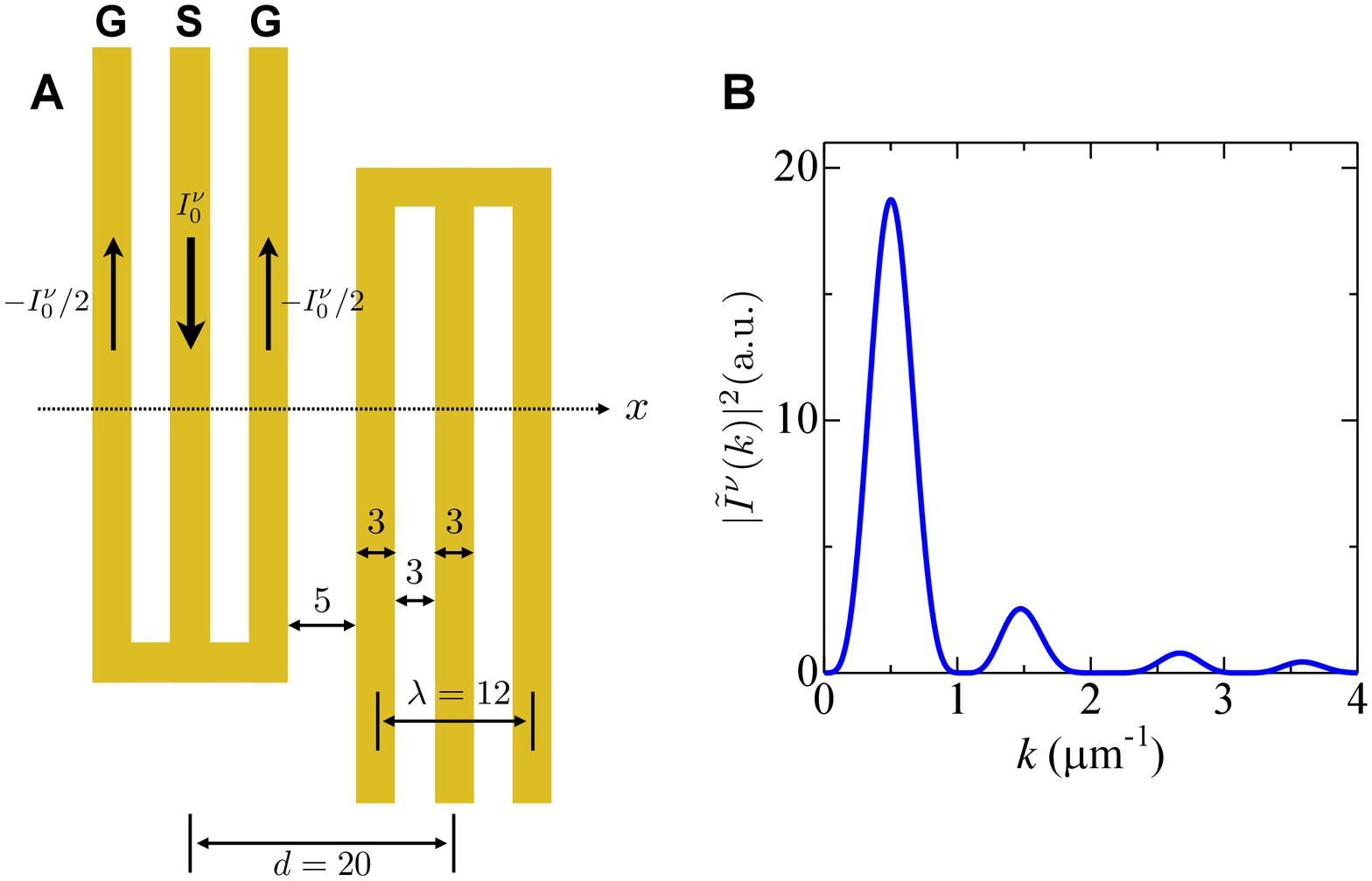}
\caption{(A) Schematic illustration of a pair of coplanar waveguides used for the spin wave spectroscopy. Each waveguide consists of one signal (S) line and two ground (G) lines. The associated current density distribution as well as length scale (in the unit of $\mu$m) are also shown. (B) The calculated wavenumber distribution of excitation current.}
\label{FigWG}
\end{center}
\end{figure}

In Fig. \ref{FigWG}A, the coplanar waveguide pattern employed in this study is illustrated. The wavelength $\lambda$ and propagation gap $d$ are 12 $\mu$m and 20 $\mu$m, respectively. Each waveguide consists of one signal line at the center and two ground lines at the both sides, which is terminated with a short circuit. When it is connected to the VNA through the GSG microprobe, the input current density for the signal and ground line is $I_0^\nu$ and -$I_0^\nu/2$, respectively. By taking the Fourier transform for the spatial distribution of current density $I^\nu (x)$, the wavenumber distribution $|\tilde{I}^\nu (k)|^2$ can be estimated\cite{AESWS_doppler_Science, AESWS_doppler_B} as shown in Fig. \ref{FigWG}B. The main peak at $k_p = 0.50 \mu$m$^{-1}$ satisfies the relationship $k_p \sim 2\pi/\lambda$, and its full width at half maximum (FWHM) is $\delta k = 0.37 \mu$m$^{-1}$. The higher order peaks are also found for the larger $k$ region, and the second largest one is at $k = 1.47 \mu$m$^{-1}$ with the amplitude 7 times smaller than that for the main peak. To simplify the discussion, we analyzed our $\Delta L_{nm}$ spectra assuming that the contribution from the main peak centered at $k_p$ is dominant.

Such a wavenumber distribution in waveguides (Fig. \ref{FigWG}B) directly affects the linewidth of ferromagnetic resonance. In Im[$\Delta L_{11}$] spectrum, the FWHM $\delta \nu$ for the resonance peak at frequency $\nu_p$ can be given as\cite{AESWS_doppler_B}
\begin{equation}
\delta \nu = \frac{v^p_g \cdot \delta k}{2\pi} + 2 \nu_p \alpha,
\label{linewidth}
\end{equation}
with $\alpha$ representing the intrinsic Gilbert damping parameter. In case of the present Cu$_2$OSeO$_3$ specimen at 740 Oe, $\delta \nu = 0.42$ GHz is obtained from the Im[$\Delta L_{11}$] spectrum of Fig. 2A in the main text. Considering the corresponding averaged spin wave group velocity $v^p_g = 6.1$ km/s taken from Fig. 3D in the main text, the first term in Eq. \ref{linewidth} gives $\sim 0.36$ GHz. This means that $\delta \nu$ mostly reflects the wave number distribution associated with the waveguide pattern. By using $\nu_p = 3.2$ GHz, we obtain the relatively small damping parameter $\alpha \sim 0.01$, which is consistent with the previous report\cite{B20_FMR}. This allows us to estimate the decay length of propagating spin wave $l_d = v^p_g/(2\pi \alpha \nu_p) = 30\mu$m\cite{AESWS_doppler_B}.

\subsection{Material properties}

Our target material, Cu$_2$OSeO$_3$ has the chiral cubic crystal lattice with space group $P2_13$\cite{COSO_Structure, COSO_Dielectric}. It contains two distinctive magnetic Cu$^{2+}$ ($S=1/2$) sites with the ratio of $3:1$, and three-up one-down type of local ferrimagnetic spin arrangement has been reported below magnetic ordering temperature $T_c \sim 59$ K\cite{COSO_Dielectric, COSO_Ferri}. This material hosts helical spin order under zero magnetic field, where spins rotate within a plane normal to the magnetic modulation vector $\vec{q}$ \cite{Cu2OSeO3_Seki, Cu2OSeO3_SANS_Seki, Cu2OSeO3_SANS_Pfleiderer}. Application of magnetic field aligns $\vec{q}$ parallel to $\vec{H}$ and turns the spin texture into the conical one. Above the critical magnitude of magnetic field $H_c$, magnetization is saturated and collinear ferrimagnetic state is stabilized. In the present study, the character of spin wave in this collinear ferrimagnetic state is mainly investigated. To simplify the analysis, we adopted the continuum approximation and treated this compound as the ferromagnetic system.

\begin{figure}
\begin{center}
\includegraphics*[width=9cm]{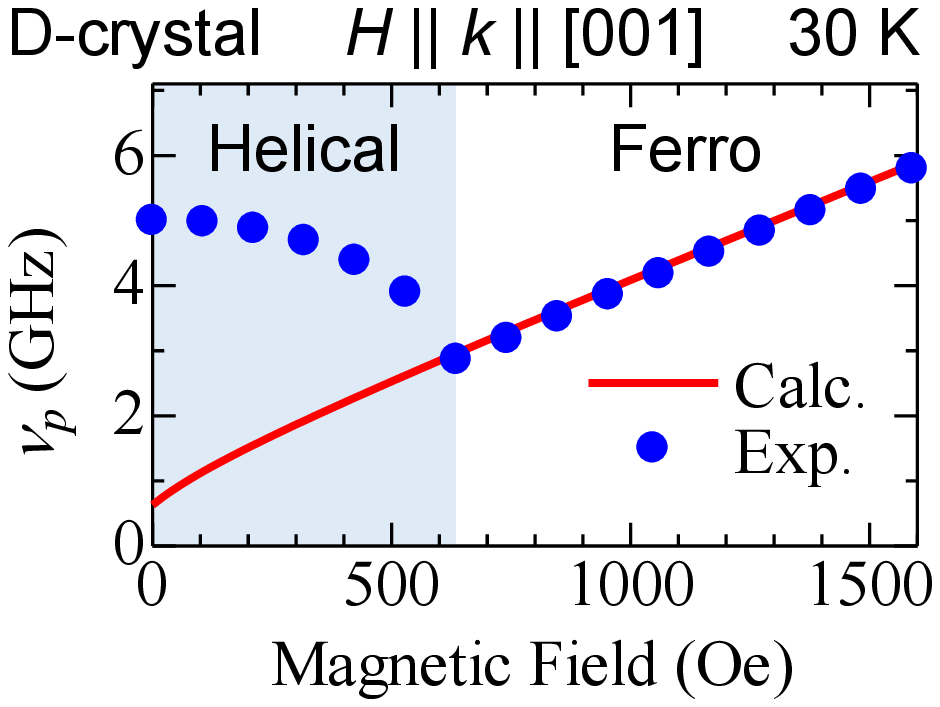}
\caption{Magnetic field dependence of resonance frequency $\nu_p$ in the $\Delta L_{21}$ spectrum, measured for the D-crystal of Cu$_2$OSeO$_3$ at 30 K. The experimental data is taken from Fig. 3A in the main text, and the theoretical fit is by Eq. 3 in the main text with $k = k_p$.}
\label{FigFit}
\end{center}
\end{figure}

The material parameters included in Eq. 3 in the main text can be estimated so as to reproduce the $H$-dependence of magnetic resonance frequency $\nu_p$ in the ferromagnetic state (Fig. \ref{FigFit}). In this process, several additional confinements are imposed\cite{Kataoka}. The helical spin modulation period $\lambda_h$ ($\sim$ 62 nm)\cite{Cu2OSeO3_SANS_Seki, Cu2OSeO3_SANS_Pfleiderer} and the corresponding magnetic wave number $Q=2\pi / \lambda_h$ in the ground state is given as
\begin{equation}
Q=-D/J,
\end{equation}
and the critical magnetic field $\mu_0 H_c$ ($\sim 0.063$ T) satisfies
\begin{equation}
\frac{\gamma \hbar}{V_0} \mu_0 H_c = \frac{D^2 S}{J} - 2K S^3.
\end{equation}
The saturation magnetization is $M_s = \hbar \gamma S/V_0 =  0.46 \mu_\textrm{B}/$Cu$^{2+}$ at 30 K, with $V_0 \sim 89$ \AA$^3$ being the volume of formula unit cell of Cu$_2$OSeO$_3$. From these restrictions, we obtain $D = 3.4 \times 10^{-4}$ J/m$^2$, $J = 3.4 \times 10^{-12}$ J/m, $K = 6.9 \times 10^{3}$ J/m$^3$, $\gamma/2\pi = 29$ GHz T$^{-1}$, $\mu_0 M_s = 0.12$ T, and $S=0.44$. These values are used to calculate the spin wave dispersion in Fig. 4, D and E in the main text and $H$-dependence of $\nu_p$ in Fig. \ref{FigFit}. Similar values have also been reported in Ref. \cite{B20_FMR}.

Note that Cu$_2$OSeO$_3$ also hosts the skyrmion spin state for narrow temperature region from 56 K to 59 K, just below $T_c$\cite{Cu2OSeO3_Seki, Cu2OSeO3_Seki, Cu2OSeO3_SANS_Pfleiderer}. Since the magnitude of propagating spin wave rapidly decays as temperature approaches $T_c$, the  experimental investigation of spin wave character in the skyrmion state remains the future challenge.

\subsection{Symmetry of Spin Current}

Spin current is generally characterized by the combination of magnetic moment $\vec{M}_0$ and wave vector $\vec{k}$. Here, $\vec{M_0}$ is an axial vector, and odd for time-reversal and even for space-inversion. In contrast, $\vec{k}$ is a polar vector, and odd for both time-reversal and space-inversion. Thus, the spin current expressed as the product of $\vec{k}$ and $\vec{M}_0$ is even for time-reversal and odd for space-inversion.

\begin{figure}
\begin{center}
\includegraphics*[width=13cm]{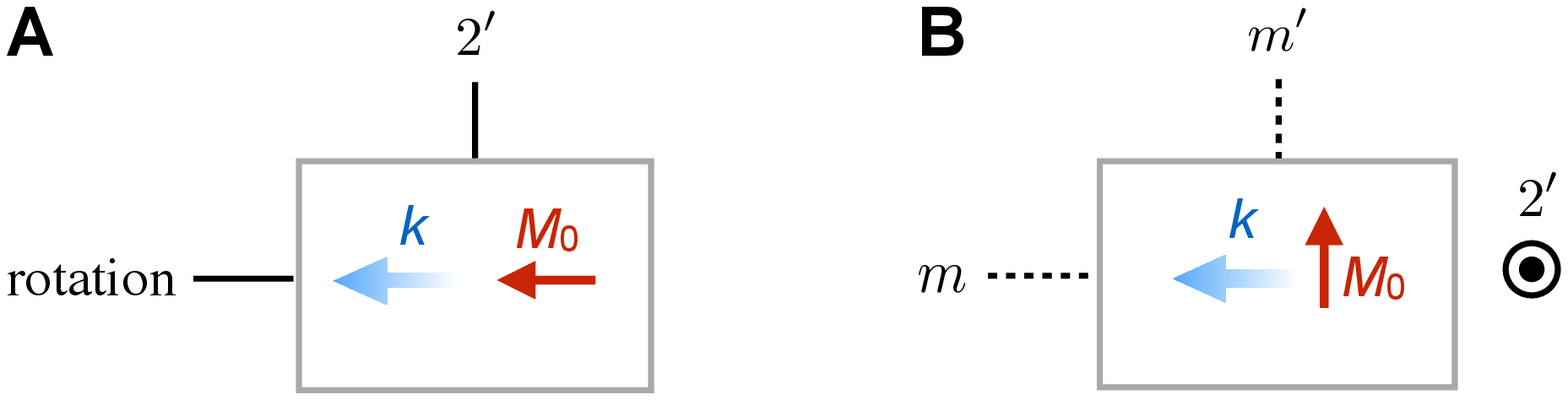}
\caption{Spin current characterized by the combination of magnetic moment $\vec{M}_0$ and wave vector $\vec{k}$, with (A) $\vec{k} \parallel \vec{M}_0$ and (B) $\vec{k} \perp \vec{M}_0$ configurations. The compatible symmetry elements are also indicated.}
\label{FigSymmetry}
\end{center}
\end{figure}

Figure \ref{FigSymmetry}, A and B summarize the compatible symmetry elements for the spin current with the $\vec{k} \parallel \vec{M}_0$ and $\vec{k} \perp \vec{M}_0$ configurations. In case of $\vec{k} \parallel \vec{M}_0$, the spin current sustains a rotation axis along the $\vec{k}$ direction and $2'$ (two-fold rotation followed by time-reversal) axis normal to it (Fig. \ref{FigSymmetry}A). Since no mirror plane or space-inversion center is present, this belongs to the chiral (and not polar) symmetry. In contrast, the spin current with the $\vec{k} \perp \vec{M}_0$ configuration has a mirror plane ($m$) normal to $\vec{M}_0$, $m'$(mirror reflection followed by time-reversal) plane normal to $\vec{k}$, and $2'$-axis normal to both $\vec{k}$ and $\vec{M}_0$ (Fig. \ref{FigSymmetry}B). This belongs to the polar (and not chiral) symmetry with the polar axis normal to both $\vec{k}$ and $\vec{M}_0$.

For each configuration, the application of space-inversion operation reverses $\vec{k}$ (but not $\vec{M}_0$), as well as the associated sign of the chirality or polarity in the spin current.

\begin{thebibliography}{100}
\bibitem{comment} The proper definition of chiral symmetry is provided in Ref. \cite{SW_Symmetry}. For the detail of symmetry analysis for spin current, also see Supporting Online Material.
\end{thebibliography}
\end{document}